\documentstyle[prl,aps,multicol]{revtex} 
\tighten 
\input epsf 
 
\newcommand{\D}{{\rm d}} 
\newcommand{\I}{{\rm i}}

\begin{document} 
\draft 
\title{Non-linear conductivity and quantum interference in  
disordered metals} 
\author{M. Leadbeater$^{(1)}$, R. Raimondi$^{(1)}$, P. Schwab$^{(2)}$,  
 and C. Castellani$^{(2)}$} 
\address{  
$^{(1)}$Istituto per la Fisica della Materia e Dipartimento di Fisica "E. Amaldi", 
 Universit\`a di Roma Tre, 
	Via della Vasca Navale 84, 00146 Roma, Italy\\  
$^{(2)}$Istituto per la Fisica della Materia e Dipartimento di Fisica, 
 Universit\`a ``La Sapienza'',  
	 piazzale A. Moro 2, 00185  Roma - Italy  
} 
\date{\today} 
\maketitle 
\begin{abstract} 
We report on a novel non-linear electric field effect in the   
conductivity of disordered conductors. We find that an electric field 
gives rise to dephasing in the particle-hole channel, which depresses 
the interference effects due to   disorder and interaction and leads to 
a non-linear conductivity. 
This non-linear effect introduces a field dependent temperature  
scale $T_E$ and provides a microscopic mechanism for electric field scaling at 
the metal-insulator transition.  
We also study the magnetic field dependence of the non-linear conductivity and 
suggest possible ways to experimentally verify our predictions.  
These effects offer a new probe to test the role of quantum 
interference at the metal-insulator transition in disordered conductors. 
 
\end{abstract} 

\begin{multicols}{2} 
Disordered conductors have been the subject  of theoretical and experimental 
study for almost twenty years\cite{altshuler85,lee85}.   
Recently there has been a 
strong resurgence of interest in the field due to the unexpected discovery of 
a metal-insulator transition  in two-dimensional systems\cite{kravchenko95}. 
Various suggestions have been made concerning  
the origin of the temperature dependence of the resistivity 
in the metallic phase and  the nature of the metal-insulator  
transition\cite{gen}. 
One main question is whether the 
transition is of a classical origin or if it is a real quantum phase 
transition.  
In the first case,  
if a standard Landau quasi-particle picture applies the observed 
resistivity could be attributed to a temperature dependent  
scattering time in the context of 
the semi-classical Boltzmann-Landau kinetic equation\cite{maslov98}.  
In the second case,  
it has been pointed out\cite{castellani98} that 
the occurrance of a metallic phase and a metal-insulator transition in two  
dimensional 
systems is indeed possible within  
the standard theory of disordered-interacting electrons 
\cite{castellani84}. 
 
To discriminate between these possibilities one needs specific probes 
for  quantum interference effects. 
Weak localization (WL) is probed effectively by the application of 
a magnetic field. This magnetic field also affects the quantum interference 
from the combined contribution of disorder  
and electron-electron interaction (EEI)\cite{altshuler80} 
but only in the particle-hole triplet channel due to the Zeeman coupling.  
In this paper we propose a new probe for the EEI contribution (both the singlet 
and triplet) based on the non-linear conductivity in the presence of a static 
electric field. We recall that WL is not affected by such a field  
\cite{altshuler85,bergmann82}. 
 
To be more specific we have found that a static (or low frequency) electric 
field: (1) acts as a source of dephasing in the particle-hole channel and  
introduces a characteristic temperature $T_E = (D_{\rm qp} e^2 
E^2)^{1\over 3}$ below which interference effects are suppressed. Here $D_{\rm 
qp}$ is the quasi-particle diffusion coefficient defined below.  (2) For 
temperatures above $T_E$ non-linear effects in the conductivity appear as 
$T_E^3/T^3$ corrections. (3) For large electric field the scale of the magnetic 
field in the magnetoresistance is set by $T_E$. Clearly one expects that besides 
quantum interference,  heating effects will also be important in the 
non-linear conductivity and we shall suggest how to distinguish the two effects. 
Our results, besides providing a new probe for EEI corrections,  offer a 
microscopic mechanism for the electric field scaling which is observed in 
two-dimensional systems\cite{kravchenko96}. 
 
Before giving details of the mathematical derivation a qualitative understanding 
of the effect may be obtained by simple physical arguments along the lines of 
Ref.\cite{aleiner98}. In a generalised Hartree-Fock picture  one electron  
is scattered by the 
potential created by all the other electrons. Due to disorder, the electron 
density is not uniform and hence this potential is random. From a semi-classical 
point-of-view, a local, single-particle quantity, like current, 
only involves closed paths. Futhermore, the  corrections are  
dominated by all the other electrons retracing backwards-in-time (as holes) 
almost the same closed paths.  
According to Ref.\cite{aleiner98}   only trajectories 
which are traversed in a time $\eta  < 1/T$  contribute to quantum corrections. 
Although the two electrons go around the same closed path they have different 
starting positions. The first electron starts at the observation point ${\bf 
x}_1$ at time zero, while the second electron will only start to retrace the 
path at the point 
of interaction ${\bf x}_2$ at time $t_1$. This means that the second electron will traverse 
part of the closed path at a different time. In the presence of a vector 
potential the accumulated phase difference is then 
$\phi_1 -\phi_2 = e \int_{t_1-\eta}^{0} \D t'   
{\dot {\bf x}}_1(t') \cdot {\bf A}({\bf x}_1(t'), t' )  
- e \int_{t_1}^{\eta} \D t'   
{\dot {\bf x}}_2(t') \cdot {\bf A}({\bf x}_2(t'), t' ).$ 
If the vector potential is  time independent 
 (eg. a magnetic field) these phases completely cancel. However, 
if the vector potential is time dependent (as for a static electric field) the 
time delay leads to a finite phase difference   
$\phi_1 -\phi_2 = e  ( {\bf x}_2 - {\bf x}_1 )\cdot {\bf E} \eta$, 
which suggests that the EEI correction should be sensitive to  a static electric 
field, in contrast to WL. Such a phase-sensitivity leads to 
 non-linear conductivity. It is possible to estimate the typical 
electric field scale where dephasing and non-linear effects in a weakly  
disordered metal become important.  
The typical time scale is the inverse temperature and the 
typical length  scale is the thermal length $L_T=(D_{\rm qp}/T)^{1/2}$. The 
non-linear effects become important when the phase difference induced by 
the electric field is of order one, 
which leads  to the condition that the voltage drop over a thermal length 
becomes of the order of the temperature, i.e., when $eEL_T \sim T$. This 
condition defines the temperature scale $T_E$ given above. 
 
We now present a quantitative theory of our results. We start  
with the expression for the EEI quantum correction to the current due to the  
interplay between disorder and interaction. Within  the real-time  
Keldysh formalism we obtain:  
\begin{eqnarray} 
\label{eq1} 
\delta {\bf j}(t) &= &- { 4 \tau^2 e \over \pi } 
\int \D\eta \D t_1 \D t_2 \left( { \pi T \over \sinh( \pi T\eta ) } \right)^2 
 \sum_q D_{\rm qp} {\bf q} \nonumber\\ 
&&\times\sum_{J,M}  
 V_{J,M} ({\bf q}, t_1-t_2) D_{J,M}^{\eta'=0}({t_2, t-\eta};{\bf q} ) 
 \nonumber\\  && \times D^{\eta}_{J,M}(t-\eta/2, t_1-\eta/2;{\bf q}) 
.\end{eqnarray} 
A pictorial representation of this equation is shown in Fig.\ref{fig1}. 
The details of its derivation may be found in \cite{raimondi99}. 
 The sum $\sum_{J,M}$  
is over one singlet ($J=0,M=0$) and three triplet channels with $J=1, M=0,  
\pm 1 $.  In eq.(\ref{eq1})  $\tau$ is the elastic scattering time  
which is the short-time cut-off in the problem.   
 $V_{J,M}$ and $D^{\eta}_{J,M}(t,t')$ are the  
interaction and the diffusion propagator in the spin channel ($J,M$).  
Here the time arguments $t$, $t'$ refer to the incoming and outgoing  
centre-of-mass time of the particle-hole pair and $\eta$ to the relative  
time which is constant during the propagation. Notice that both $V_{J,M}$  
and $D_{J,M}$ are retarded functions. The factor containing the hyperbolic  
sine comes from the Fourier transform of Fermi functions and limits us to 
trajectories with traverse time $\eta<1/T$. The interaction  
is found by summing ladder diagrams and is given by 
\begin{eqnarray} 
\label{eq3} 
V_{J, M}({\bf q}, \omega) &=&  { \gamma_J  } 
{- \I \omega +D_{\rm qp}q^2 + \I M \tilde \Omega_s \over 
 - \I(1-2\gamma_J ) \omega +D_{\rm qp}q^2 +\I M \tilde \Omega_s } 
,\end{eqnarray} 
where $\gamma_J$ is the static amplitude for scattering between  
particles and holes.  The quasi-particle diffusion constant  
can be expressed in terms of the particle diffusion constant $D_{\rm qp}=D/Z$.  
The parameter $Z$, which arises in the context of the Fermi  
liquid theory of disordered systems \cite{castellani84} as the energy  
renormalization,  plays the role of mass renormalization,  $m^{*}/m$, 
in the effective Fermi liquid theory of disordered systems 
\cite{castellani87}. 
The interaction amplitude in  
the spin singlet channel is given by $\gamma_{J=0} = 1/2$ for long range  
Coulomb forces.  
The triplet amplitude, for which we adopt in the following the standard  
notation 
$\gamma_{J=1}=-\gamma_2/2$,  is related to 
the Landau parameter $F_a^0$ via $\gamma_2=-A^{0}_{a}=-F_a^0/(1+F_a^o)$.  
The diffusion propagator $D_{J,M}$ is given 
by the solution of the differential equation 
\begin{eqnarray} 
\label{eq2} 
&\left\{ {\partial \over \partial t} 
+ {D_{\rm qp}} \left[-\I\nabla +e{\bf A}_{\eta}({\bf r},t)  
\right]^2 + \I M \tilde \Omega_s  \right\} D^{\eta}_{J,M}(t,t')  &\nonumber\\ 
& = {1\over {\tau} }\delta(t-t') \delta({\bf r}- {\bf r'}).& 
\end{eqnarray} 
where  
${\bf A}_{\eta}({\bf r},t)={\bf A}({\bf r},t+\eta/2) 
 - {\bf A}({\bf r},t-\eta/2)$. 
The term $\I M\tilde{\Omega}_s$ is  
due to the Zeeman coupling,  
where $\tilde \Omega_s = (1+\gamma_2) \Omega_s $ with  
$\Omega_s = g \mu_B H$. 
  
For a better understanding of  
eq.(\ref{eq1}) we make contact with the physical arguments 
given in the introduction.  
By considering Fig.\ref{fig1} one  observes that  
for a piece of the path the particle and hole are delayed by a time $\eta$, the 
traverse time of the closed path.  
This  
corresponds to the second of the diffusons in eq.(\ref{eq1}). In the other 
piece there is no delay between the particle and hole. This corresponds 
to the first diffuson in eq.(\ref{eq1}).  
 
We now evaluate the current explicitely. According to eq.(\ref{eq2}) the  
interaction $V_{J,M}$ and the first of the two diffusons in (\ref{eq1}) do not  
depend on the vector potential. An electric field, however, affects the  
second diffuson in (\ref{eq1}) due to the non-zero time delay $\eta$  
between the particle and hole. For a static field the vector potential  
is ${\bf A}(t) = -{\bf E} t $ and  
the solution of  (\ref{eq2}) is 
$ 
D^{\eta}_{J,M}\left(t-{\eta\over 2}, t_1-{\eta\over 2}; {\bf q} \right) 
\nonumber\\ =  
{1\over {\tau}} \exp \left\{  
-\left[ {D_{\rm qp}}( {\bf q}-e {\bf E} \eta )^2  
-\I M \tilde \Omega_s \right](t-t_1) \right\}$. 
The equation for the current after integrating 
over the momentum then becomes 
\begin{eqnarray} 
\delta {\bf j}_{J, M }& =&- {\bf E}{4 e^2D_{\rm qp}  \over  \pi}  
\gamma_J\left( { 1-2 \gamma_J} \over 4 \pi D_{\rm qp} \right)^{d/2} \nonumber\\ 
&&\int_\tau^\infty { \D\eta }  
\left[ { \pi T  \over \sinh( \pi T \eta ) } \right]^2 
\int_0^\eta \D t_1 
{ t_1 \eta \over (\eta -2 \gamma_J t_1 )^{1+d/2}} \nonumber\\ 
&&\times \cos [M \Omega_s ( \eta -2\gamma_J t_1 ) ]\nonumber\\ 
&& \times \exp \left[- 
T_E^3 \eta^2 t_1(\eta-t_1)/(\eta-2 \gamma_J t_1) 
\right]  
\label{eqdj} 
\end{eqnarray} 
where we have introduced  $T_E^3 = D_{\rm qp}(eE)^2$ and $d$ is the
dimension. 
It is clear from this equation that the electric field provides 
a dephasing time $\sim T_E^{-1}$, since in the low temperature limit 
$T\ll T_E$ the exponential now cuts off all times larger than $T_E^{-1}$. 
 
We first consider the current in the weak electric field regime  and derive 
the leading non-linear terms. 
In the absence of magnetic field  we find  
\begin{eqnarray} 
\delta {\bf j} &=& {\bf E} 
\frac{e^2}{2^{d-1}\pi^{2}} 
\left(\frac{D_{\rm qp}}{ T}\right)^{2-d\over 2} 
\int^\infty_{\pi T \tau} d x  
\frac{x^{2-\frac{d}{2}}}{\sinh^2 (x)} \nonumber\\ 
&&\times \left( f^1_d(\gamma_2)  
+ f^3_d(\gamma_2 )   
\frac{x^3 T_E^3}{ \pi^3 T^3} \right)\end{eqnarray}  
where the functions $f_d^{1,3}$ are shown in the table. 
For the sake of completeness we have also included  
the term  linear in the electric field which reproduces 
 the well-known interaction correction to the conductivity.  
Notice that the functions $f_d^{1,3}(\gamma_2 )$ are the sum of 
the singlet and the triplet contributions.  
The remaining integrals are of a standard form.
We now have the results for $\delta \sigma = \delta |{\bf j}|/|{\bf E}|$ 
\begin{eqnarray} 
\delta \sigma_1  &=&  \frac{e^2}{\pi^{2}} L_T 
\left[ -2.46 f^1_1(\gamma_2)  
+\frac{ 4.88}{\pi^3} f^3_1(\gamma_2) \frac{ T_E^3}{T^3} 
\right] 
\cr  
\delta \sigma_2 &=&  
\frac{e^2}{2\pi^{2}} 
\left[ -f^1_2(\gamma_2) 
\ln\left(\frac{\rm e}{2 \pi T\tau}\right) 
+\frac{\pi}{30} f^3_2(\gamma_2) \frac{T_E^3}{T^3} 
\right] \cr 
\delta \sigma_3 &=& 
 \frac{e^2}{4\pi^{2}} 
L_T^{-1}\left[1.83 f^1_3(\gamma_2) 
+ \frac{ 2.32}{\pi^3} f^3_3(\gamma_2) \frac{T_E^3}{ T^3 } 
\right]  
\end{eqnarray} 
where $L_T$ is the thermal length defined previously and we have left out 
temperature independent terms. We recall that, 
in the case of spin-singlet interactions only ($\gamma_2 = 0$) 
the conductivity decreases with decreasing temperature (i.e. $f^1_d(0)>0$),  
whereas for sufficiently 
large triplet amplitude 
$\gamma_2$ the latter dominates and leads to an increase of conductivity 
with decreasing temperature (i.e. $f^1_d(\gamma_2)<0$). 
The non-linear coefficient $f_d^3 (\gamma_2 )$, however, is generically positive 
and only  
changes sign for large $\gamma_2$ in $d=3$. 
 
We study  
the cross-over behaviour from small to large electric fields  
by numerically integrating eq.(\ref{eqdj}). The conductivity as a function of 
the electric field is plotted in fig.\ref{fig2} for two values of $\gamma_2$ 
for $d=2$. 
At zero field, for $\gamma_2 =0$ ($\gamma_2 =5$) the correction is localising 
(anti-localising) with $\delta \sigma_2 < 0$ ($\delta \sigma_2 > 0 $). 
The quadratic increase at small fields has a positive curvature irrespective 
of the value of $\gamma_2$. At large field, the temperature scale disappears 
and the correction $\delta \sigma$ has the same form as  the linear 
conductivity with $T_E$ replacing the temperature. 
 
Non-linear effects also appear in the magnetoconductance which originate from 
the magnetic field depression of the $M=\pm 1$ 
triplet contributions to the current.  
In particular we find for small $T_E$ and small Zeeman energy $\Omega_s$ 
\begin{eqnarray} 
\Delta \sigma_1 &=& 
-  \frac{e^2}{\pi^2}  
L_T \frac{\Omega_s^2}{T^2} 
\left[ 
\frac{2.32}{\pi^2} g^1_1(\gamma_2) 
+\frac{41.85}{\pi^5}g^3_1(\gamma_2)\frac{T_E^3}{T^3} \right] \cr 
\Delta \sigma_2 &=&  
-\frac{e^2}{2 \pi^2} \frac{\Omega_s^2}{T^2}  
\left[  \frac{3\zeta(3)}{2\pi^2} g^1_2(\gamma_2) + 
        \frac{\pi}{42}           g^3_2(\gamma_2) \frac{T_E^3}{T^3} 
 \right] \cr 
\Delta \sigma_3 &=&  
- \frac{e^2}{4\pi^2}L_T^{-1} 
 \frac{\Omega_s^2}{T^2}  
\left[  
\frac{1.58}{\pi^2} g^1_3(\gamma_2) + 
\frac{13.04}{\pi^5}g^3_3(\gamma_2) \frac{T_E^3}{T^3} 
\right]  
\end{eqnarray} 
where $\Delta \sigma=\sigma (\Omega_s) - \sigma(0)$ and  
the functions $g_d^{1,3}$ are also shown in the table\cite{note}.  
To illustrate the behaviour at large $\Omega_s$ we again resort to numerical 
integration of eq.(\ref{eqdj}). In fig.\ref{fig3}  we show the 
magnetic field dependence of the current for two choices of $T_E$ for  
$\gamma_2 = 2.5$. Notice that for such a value of $\gamma_2$ the zero  
magnetic field conductivity interference correction has an anti-localizing  
character. This 
explains the suppression of conductivity with increasing $T_E$.  
For small $T_E$ we obtained the standard behaviour of a initial  
quadratic decrease on the scale of the temperature followed by a  
logarithmic suppression of the corrections. For large $T_E$ however,  
the temperature disappears as an energy scale and, although the curve  
appears similar, the scale of the magnetic field is now set by $T_E$.  
Expanding eq.(\ref{eqdj}) 
to leading order in the magnetic field for $T_E\gg T$ one obtains  
$\Delta \sigma \propto \Omega^2_s/T_E^2$.  

The effects described in this paper may be detected by measuring the  
current-voltage characteristics.  
In such a measurement however the electron temperature changes  
with the applied voltage and one has  
to discriminate heating from non-heating non-linear effects. A direct way to 
isolate the non-linear contribution due to the dephasing effect of $E$  
would be to measure the electron temperature $T_{el}$ for a given $E$ (for 
instance by noise measurements as in \cite{shotnoise}). Then 
$\sigma(T_{el},0)-\sigma(T,E)$ yields the effect of the electric field on the 
EEI contribution and provides a direct probe of the relevance of quantum 
interference in the p-h channels. Alternatively, at 
low temperature, where $T_{\rm el} \tau_{\rm el-ph } \gg 1$  
($\tau_{\rm el-ph}$ is the heat electron-phonon relaxation time) non-heating  
effects could be detected by exploiting the different time scales  
$\tau_{\rm el-ph }$ and $T_{el}^{-1}$ which control the frequency  
dependence of heating and non-heating effects respectively.  
In a time-dependent electric field, $E(t)= E \cos( \Omega t )$, the electron  
temperature becomes time dependent. For frequencies  
$\Omega > 1/\tau_{\rm el-ph}$ however the temperature cannot follow the  
electric field, i.e. heating becomes time independent. Non-linearities  
due to quantum interference on the other hand follow the electric field  
instantly as long as the frequency remains smaller than the temperature.  
Thus measuring non-linear response in the presence of a microwave with a frequency  
of the order $\Omega \geq 1/\tau_{\rm el-ph}$ offers  a possibility to  
detect the predicted effects\cite{suggestion}. 
 
Possible non-heating effects have already  
been observed in different materials like 
2D Si-MOSFETs\cite{kravchenko96,vitkalov88}, 
GaAs\cite{yoon99}, SiGe\cite{senz99} and in gold films\cite{bergmann90,liu91}.  
In particular in Ref.\cite{kravchenko96} a remarkable electric field scaling 
was observed near the metal-insulator transition. Our theory of the  
non-linear effects provides an explicit mechanism for the electric field  
scaling via the temperature scale $T_E$: 
on the basis 
of general scaling arguments close to a quantum critical point, 
the temperature scales as $T\sim \xi^{-z}$ where $\xi$ is the correlation 
length and $z$ is the dynamical critical exponent\cite{sondhi97}.  
In a diffusive system  
temperature and length scales are related via the diffusion constant with  
$T \sim D_{\rm qp}(\xi)/ \xi^2$ implying a scaling of $D/Z = D_{\rm qp}$  
near the critical point as $D_{\rm qp}\sim \xi^{2-z}$.  From our relation 
$T^3_E=D_{\rm qp}e^2E^2$  
one then obtains $E\sim \xi^{-(1+z)}$.  
In the experiments $z$ is near one, which  
corresponds to a growing quasi-particle diffusion constant and a vanishing  
quasi-particle density of states near the transition. This small value of $z<2$ 
implies that the electronic specific heat 
would vanish as $c_v\sim T\xi^{z-2}\sim 
T^{2/ z}$. From these  
considerations one expects large non-linear effects near the critical point. 
Such large effects have been   observed  
in Ref.\cite{yoon99} 
for a GaAs metallic sample near the metal-insulator transition. 
In this experiment the differential conductivity was found to decrease 
with increasing voltage.  
In particular at low temperature the conductivity shows  
a non-monotonic behaviour 
of the type shown in fig.2 for large $\gamma_2$.  
By comparing with experiment it appears that the scale over which 
the effect is observed  is 
larger than what we would predict based upon a naive estimate of the diffusion 
constant from the conductivity $\sigma = 2 e^2 D N_0$  
and the free particle density of states $N_0$ (i.e. assuming $D_{\rm qp}\sim D$). 
However, by allowing for quasi-particle diffusion constant renormalization as 
implied by scaling one can obtain larger effects.

 
Finally, we point out that the effects discussed in this paper could have
an enhanced relevance  in the presence of strong local electric fields 
such as in percolative metallic  systems. 
 
This work was partially supported by MURST under contract no. 9702265437  
(R.R. and C.C.), by INFM under the PRA-project QTMD (R.R.) and the European  
Union TMR program (M.L. and P.S.). C.C. and R.R. acknowledge useful  
discussions with M. Sarachik and S. Vitkalov.  

{\centerline{\epsfxsize=5cm\epsfbox{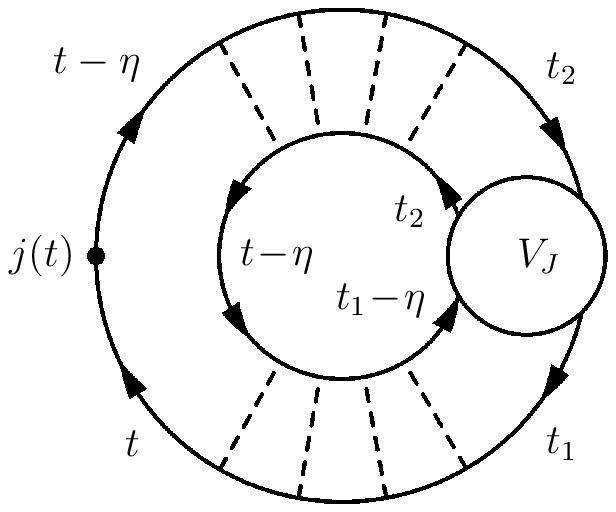}}} 
\begin{figure} 
\narrowtext 
\caption{Pictorial representation of the current formula. Four dashed lines 
represent a diffuson.} 
\label{fig1} 
\end{figure} 
 
{\centerline{\epsfxsize=5cm\epsfbox{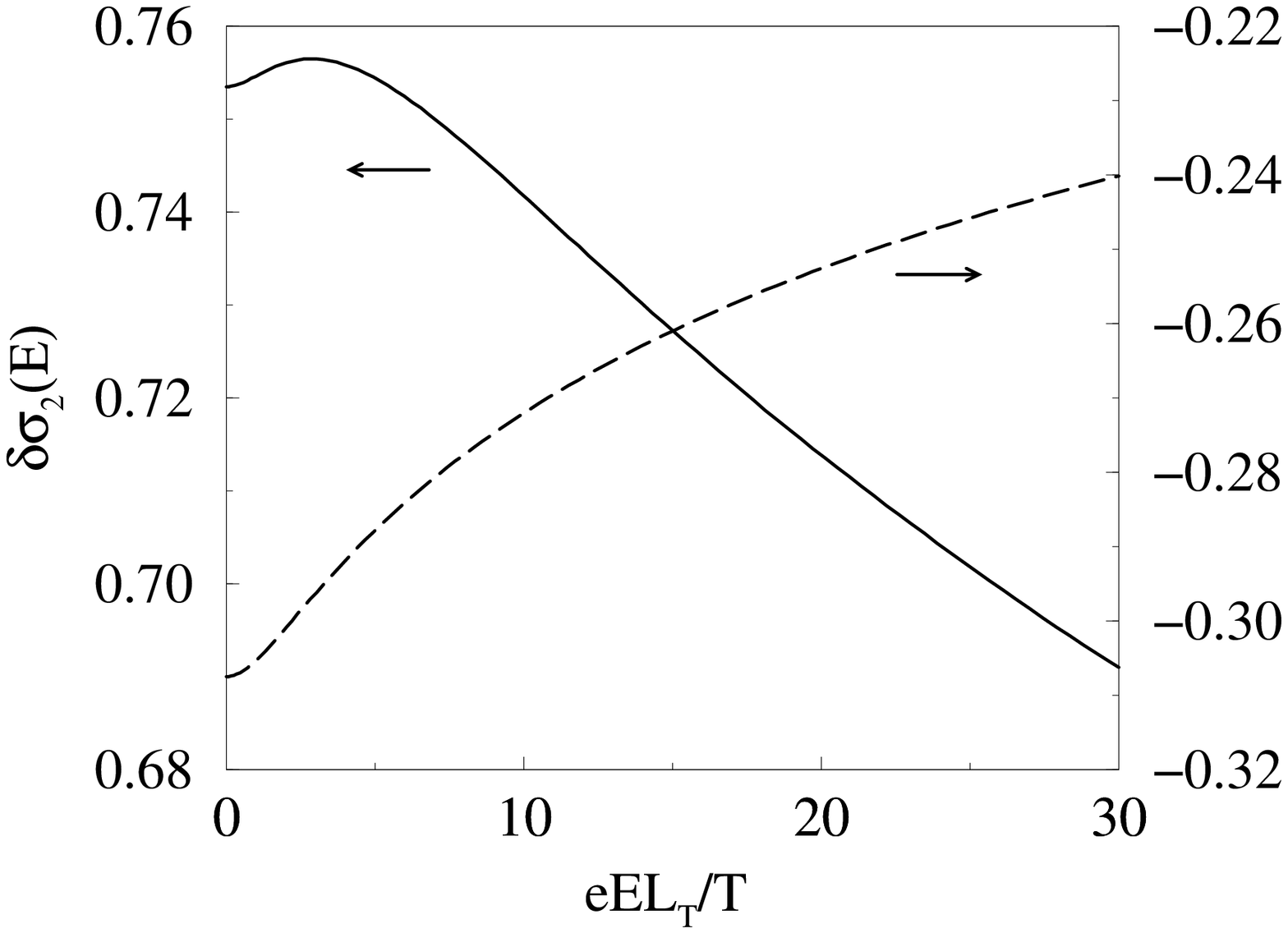}}} 
\begin{figure}  
\narrowtext 
\caption{The electric field dependence of the interaction correction to the 
conductivity in two 
dimensions in units of $e^2/\hbar$ for $\gamma_2=0$ (dashed line) 
 and $\gamma_2=5$ (solid line). Note the different scales used for the two 
 values of $\gamma_2$.} 
\label{fig2} 
\end{figure}

{\centerline{\epsfxsize=5cm\epsfbox{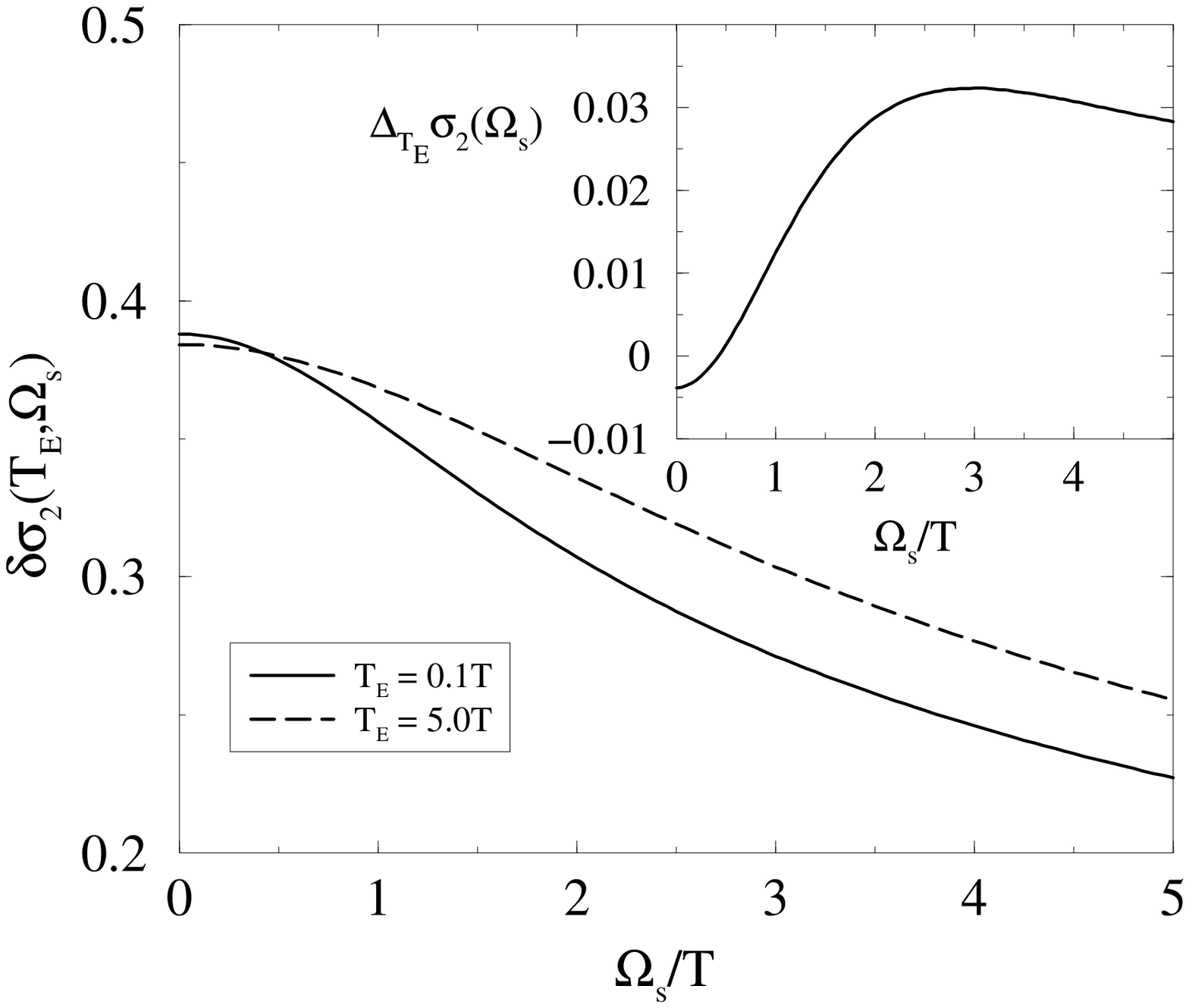}}} 
\begin{figure} 
\narrowtext 
\caption{The magnetic field dependence of the  
conductivity in two dimensions in units of $e^2/\hbar$ 
for two different values of $T_E$ and for $\gamma_2=2.5$. The inset 
shows the difference $\Delta_{T_E}\sigma_2 (\Omega_s )=\delta\sigma_2 (5T ,\Omega_s )- 
\delta\sigma_2 (0.1T ,\Omega_s )$of the two curves.} 
\label{fig3} 
\end{figure} 
 
\begin{table} 
\narrowtext 
\begin{tabular}[]{|c||c|} 
$f^1_d(\gamma_2)$ & $\frac{2}{d} - 3  
\frac{4(1+\gamma_2)^{\frac{d}{2}}-4-2 d \gamma_2} 
{(d-2) d \gamma_2}$ \\\hline 
$f^3_d(\gamma_2)$ & $ \frac{4}{d(d+2)} \left( 1+ 
 3 \frac{[24+(16-4d)\gamma_2](1+\gamma_2)^{\frac{d}{2}} 
-24-(2+d)\gamma_2(8+d \gamma_2)} 
{(d-4)(d-2)\gamma_2^3}\right)$ \\ \hline 
$f_2^1(\gamma_2 )$ & $1+3\left[1 - {1+\gamma_2 \over \gamma_2 }\ln ( 
 1+\gamma_2 )  
\right] $\\ \hline     
$f_2^3(\gamma_2 )$ & $\frac{1}{2}+\frac{3}{2}  
\left[ { 6+ 5 \gamma_2 \over \gamma_2^2 } -  
{(6+2\gamma_2)(1+\gamma_2) \over \gamma_2^3 }\ln(1+\gamma_2 ) \right]$ \\ \hline 
$g^1_d(\gamma_2)$ & $2 \frac{2(1+\gamma_2)^{\frac{d}{2}}-(1+\gamma_2)^2 
[2+(d-4) \gamma_2]}{(d-6)(d-4)\gamma_2} $\\ \hline 
$g^3_d(\gamma_2) $ & $ 4\frac{[24+4(8-d)\gamma_2](1+\gamma_2)^{\frac{d}{2}} 
-\{24+(d-2)\gamma_2[8+(d-4) \gamma_2]\} (1+\gamma_2)^{2}} 
{(d-8)(d-6)(d-4)(d-2)\gamma_2^3}$ \\ \hline  
  $ g_2^1(\gamma_2 ) $ &   
$ \frac{1}{2}\gamma_2(1+\gamma_2)$ \\ \hline 
 $g_2^3 (\gamma_2 )$ & $ \frac{ 1+\gamma_2  }{\gamma_2^2} 
\left[{3\gamma_2 + 6 -\gamma_2^2\over 6}- 
   {1+\gamma_2\over \gamma_2}\ln(1+\gamma_2) \right] $\\  
\end{tabular} 
\caption{Table of coefficients which appear in the expressions  
for the current. For small $\gamma_2$ these reduce to 
$f^1_d(\gamma_2)=2/d-3\gamma_2/2$,  
$f^3_d(\gamma_2)=4/(d(2 + d))-\gamma_2/4$, 
$g^1_d(\gamma_2)=\gamma_2/2$ and 
$g^3_d(\gamma_2)=-\gamma_2/12$.} 
\label{table1} 
\end{table} 
 
\end{multicols}


\begin{references} 
\bibitem{altshuler85} 
                B.L. Altshuler, A.G. Aronov, D.E. Khmelnitskii, and A.G. Larkin, 
                in {\em Quantum Theory of Solids}, edited by I.M. Lifshitz 
                (MIR Publishers, Moscow, 1982); 
                B.L. Altshuler and A.G. Aronov,  
                in {\em Electron-Electron Interactions in Disordered Systems}, 
                edited by M. Pollak and A.L. Efros (North-Holland,  
                Amsterdam, 1985), p. 1. 
\bibitem{lee85} P.A. Lee and T.V. Ramakrishnan, 
                Rev. Mod. Phys. {\bf 57}, 287 (1985); 
		D. Belitz and T.R. Kirkpatrick, 
		Rev. Mod. Phys. {\bf 66}, 261 (1994). 
\bibitem{kravchenko95} S.V. Kravchenko, W.E. Mason, G.E. Bower, J.E. Furneaux, 
                V.M. Pudalov, and M. D'Iorio   Phys. Rev. B \textbf{51}, 
                7038 (1995). 
\bibitem{gen}   V. Dobrosavljevic, E. Abrahams, E. Miranda, and  
                S. Chakravarty  Phys. Rev. Lett. \textbf{79},(1997) 455; P. 
		Phillips {\it et al},  Nature {\bf 395}, 253, 1998;  
		S.Chakravarty {\it et al} 
		cond-mat/9805383; V.M. Pudalov JETP Lett., {\bf 66}, 175 (1997);  
		D. Belitz and  
		T.R. Kirkpatrick, Phys. Rev. {\bf B58}, 8214-8217 (1998);  
		Q. Si and C. M. Varma, Phys. Rev. Lett. 81, 4951 (1998). 
\bibitem{maslov98} B.L. Altshuler and D.L. Maslov,  
                Phys. Rev. Lett. {\bf 82}, 145 (1999); T.M. Klapwijk and S.  
		Das Sarma cond-mat/9810349; S. Das Sarma and E.H. Hwang 
		cond-mat/9812216. 
\bibitem{castellani98}C. Castellani, C. Di Castro, and P.A. Lee, 
		Phys. Rev. B {\bf 57}, R9381 (1998). 
\bibitem{castellani84}A.M. Finkel'stein JETP {\bf 57}, 97 (1983); 
                C. Castellani, C. Di Castro, P.A. Lee, and 
                M. Ma, Phys. Rev. B {\bf 30}, 527 (1984). 
\bibitem{altshuler80}B.L. Altshuler, A.G. Aronov, and P.A. Lee, 
                Phys. Rev. Lett. {\bf 44}, 1288 (1980); 
		B.L. Altshuler, D. Khmel'nitzkii, A.I. Larkin, and P.A. Lee, 
		Phys. Rev. B {\bf 22}, 5141 (1980). 
\bibitem{bergmann82}G. Bergmann, Z. Phys.B {\bf 49}, 133 (1982). 
\bibitem{kravchenko96}S.V. Kravchenko, D. Simonian, M.P. Sarachik, 
                W. Meson, and G.E. Fourneaux, 
		Phys. Rev. Lett. {\bf 77}, 4938 (1996), 
                R. Heemskerk and T.M. Klapwijk, 
                Phys. Rev. B {\bf 58}, R1754 (1998). 
\bibitem{aleiner98} I.L. Aleiner, B.L. Altshuler and E. Gersherson, 
                cond-mat/9808053.		 
\bibitem{bergmann87}G. Bergmann, Phys. Rev. B {\bf 35}, 4205 (1987). 
\bibitem{raimondi99}R. Raimondi, P. Schwab, and C. Castellani, 
                cond-mat/9903146.  
\bibitem{castellani87} C. Castellani and C. Di Castro, 
                Phys. Rev. {\bf B34}, 5935 (1986); C. Castellani, G. Kotliar, 
		and P.A. Lee  Phys. Rev. Lett. {\bf59}, 323 (1987). 
\bibitem{note}  Note that the numerical factor we obtain in front of the 
                $g^1_2(\gamma_2)$ term 
                is slightly larger (about 10\%)  
                than the factor given in the literature.  
\bibitem{shotnoise} R. de-Picciotto {\it et al}, Nature {\bf 389}, 162 (1997). 
\bibitem{suggestion} S.A. Vitkalov (Private communication).		 
\bibitem{vitkalov88}S.A. Vitkalov, G.M. Gusev, Z.D. Kvon, G.I. Leviev, and  
                V.I. Fal'ko, 
                Sov. Phys. JETP {\bf 67}, 1080 (1988). 
\bibitem{yoon99}J. Yoon, C.C. Li, D. Shahar, D.C. Tsui, and M. Shayegan, 
		Phys. Rev. Lett. {\bf 82}, 1744 (1999). 
\bibitem{senz99}V. Senz, et al, preprint, cond-mat/9903367.  
\bibitem{bergmann90}G. Bergmann, Wei Wei, Yao Zou, and R.M. Mueller, 
                Phys. Rev. B {\bf 41} 7386 (1990). 
\bibitem{liu91} J. Liu and N. Giordano, Phys. Rev. B {\bf 43}, 1385 (1991). 
\bibitem{sondhi97} For a recent review see 
                S.L. Sondhi, S.M. Girvin, J.P Carini, and D. Shahar,   
                Rev. Mod. Phys. {\bf 69}, 315 (1997). 
\end{references}
\end{document}